\documentclass[%
 reprint,
 amsmath,amssymb,amsthm,
 nofootinbib,
 aps,
]{revtex4-1}

\usepackage{dcolumn}
\usepackage{bm}
\usepackage[final]{changes}
\usepackage{graphicx}
\usepackage{csquotes}

\begin{document}
\preprint{APS/123-QED}

\title{Quantum Mechanics from Relational Properties\\ Part III: Path Integral Implementation }

\author{Jianhao M. Yang}
\email{jianhao.yang@alumni.utoronto.ca}
\affiliation{
San Diego, CA 92121, USA
}

\date{\today}

\begin{abstract}
Relational formulation of quantum mechanics is based on the idea that relational properties among quantum systems, instead of the independent properties of a quantum system, are the most fundamental elements to construct quantum mechanics. In a recent paper (J. M. Yang, Sci. Rep. 8:13305, 2018), basic relational quantum mechanics framework is formulated to derive quantum probability, Born's Rule, Schr\"{o}dinger Equations, and measurement theory. This paper \added{further extends the reformulation effort in three aspects. First, it gives a clearer explanation of the key concepts behind the framework to calculate measurement probability. Second, we} provide a concrete implementation of the relational probability amplitude by extending the path integral formulation. The implementation not only clarifies the physical meaning of the relational probability amplitude, but also \replaced{allows us to elegantly explain the double slit experiment, }{gives several important applications. For instance, the double slit experiment can be elegantly explained. A path integral representation of the reduced density matrix of the observed system can be derived. Such representation is shown valuable} to describe the interaction history between the measured system and a series of measuring systems, \replaced{and }{. More interestingly, it allows us to develop a method }to calculate entanglement entropy based on path integral and influence functional. \replaced{In return, the implementation brings back new insight to path integral itself by completing the explanation on why measurement probability can be calculated as modulus square of probability amplitude. Lastly, we clarify the connection between our reformulation and the quantum reference frame theory. A complete relational formulation of quantum mechanics needs to combine the present works with the quantum reference frame theory.}{Criteria of entanglement is proposed based on the properties of influence functional, which may be used to determine entanglement due to interaction between a quantum system and a classical field.}
\begin{description}
\item[Keywords] Relational Quantum mechanics, Path Integral, Entropy, Influence Functional,\\ Quantum Reference Frame
\end{description}
\end{abstract}
\maketitle

\section{Introduction}
\label{intro}
Quantum mechanics was originally developed as a physical theory to explain the experimental observations of a quantum system in a measurement. In the early days of quantum mechanics, Bohr had emphasized that the description of a quantum system depends on the measuring apparatus~\cite{Bohr, Bohr35, Jammer74}. In more recent development of quantum interpretations, the dependency of a quantum state on a reference system was further recognized. The relative state formulation of quantum mechanics~\cite{Everett, Wheeler57, DeWitt70} asserts that a quantum state of a subsystem is only meaningful relative to a given state of the rest of the system. Similarly, in developing the theory of decoherence induced by environment~\cite{Zurek82, Zurek03, Schlosshauer04}, it is concluded that correlation information between two quantum systems is more basic than the properties of the quantum systems themselves. Relational Quantum Mechanics (RQM) further suggests that a quantum system should be described relative to another system. There is no absolute state for a quantum system~\cite{Rovelli96, Rovelli07}. Quantum theory does not describe the independent properties of a quantum system. Instead, it describes the relation among quantum systems, and how correlation is established through physical interaction during measurement. The reality of a quantum system is only meaningful in the context of measurement by another system. \added{The RQM idea was also interpreted as perspectivalism in the context of modal interpretation of quantum mechanics~\cite{Bene}, where properties are assigned to a physical system from the perspective of a reference system.}

\added{The idea of RQM is thought provoking. It essentially implies two aspects of relativity. The first aspect of RQM is to insist that a quantum system must be described relative to a reference system. The reference system is arbitrarily selected. It can be an apparatus in a measurement setup, or another system in the environment. A quantum system can be described differently relative to different reference systems. The reference system itself is also a quantum system, which is called a quantum reference frame (QRF). There are extensive research activities on QRF, particularly how to ensure consistent descriptions when switching QRFs~\cite{QRF1, QRF2, QRF3, QRF4, QRF5, QRF6, QRF7, QRF8, QRF9, QRF10, QRF11, QRF12, QRF13, QRF14, QRF15, QRF16, Brukner, Hoehn2018}. Noticeably, Ref.~\cite{Brukner, Hoehn2018, Hoehn2019, Yang2020, Ballesteros} completely abandon any external reference system and the concept of absolute state. Physical description is constructed using relational variables from the very beginning within the framework of traditional quantum mechanics. In addition, all reference systems are treated as quantum systems instead of some kinds of abstract entities. Treating a reference frame as a classical system, such as how the relativity theory does, should be considered as an approximation of a more fundamental theory that is based on QRF.

The second aspect of RQM is more fundamental. Since the relational properties between two quantum systems are considered more basic than the independent properties of one system, the relational properties, instead of the independent properties, of quantum systems should be considered as a starting point for constructing the formulation of quantum mechanics itself. Questions associated with this aspect of RQM include how to quantify the relational properties between two quantum systems, and how to reconstruct a quantum mechanics theory from relational properties. Note that the relational properties themselves are relative to a QRF. Different observers can ascribe a quantum system with different sets of relational properties relative to their choices of QRFs. 

It is this second aspect of RQM that inspires our works here. The effort to reconstruct quantum mechanics itself from relational properties was initiated in the original RQM paper~\cite{Rovelli96} and had some successes, for example, in deriving the Schr\"{o}dinger Equation. This reconstruction is based on quantum logic approach. Alternative reconstruction that follows the RQM principle but based on information theory is also developed~\cite{Hoehn2014, Hoehn2015}. These reconstructions appear rather abstract, not closely connect to the physical process of a quantum measurement. We believe that the relational properties should be identified in a measurement event given the idea that the reality of a quantum system is only meaningful in the context of measurement by another system. With this motivation, recently we proposed a formulation~\cite{Yang2017} that is based on a detailed analysis of quantum measurement process. What is novel in our formulation is a new framework for calculating the probability of an outcome when measuring a quantum system, which we briefly describe here. In searching for the appropriate relational properties as the starting elements for the reconstruction, we recognize that a physical measurement is a probe-response interaction process between the measured system and the measuring apparatus. This important aspect of measurement process, although well-known, seems being overlooked in other reconstruction efforts. Our framework for calculating the probability, on the other hand, explicitly models this bidirectional process faithfully. As such, the probability can be derived from product of two quantities and each quantity is associated with a unidirectional process. The probability of a measurement outcome is proportional to the summation of probability amplitude product from all alternative measurement configurations. The properties of quantum systems, such as superposition and entanglement, are manifested through the rules of counting the alternatives. As a result, the framework gives mathematically equivalent formulation to Born's rule. Wave function is found to be summation of relational probability amplitudes, and Schr\"{o}dinger equation is derived when there is no entanglement in the relational probability amplitude matrix. Although the relational probability amplitude is the most basic properties, there are mathematical tools such as wave function and reduced density matrix that describe the observed system without explicitly called out the reference system. Thus, the formulation in Ref.~\cite{Yang2017} is mathematically compatible to the traditional quantum mechanics. We restrict our analysis on ideal projective measurement in order to focus on the key ideas, with the expectation that the framework based on an ideal measurement can be extended to more complex measurement theory in future researches. }

\deleted{The idea that relational properties are more basic than the independent properties of a quantum system is profound. It should be considered a starting point for constructing the formulation of quantum mechanics. However, traditional quantum mechanics always starts with an observer-independent quantum state. It is of interest to see if a quantum theory constructed based on relational properties can address some of the unanswered fundamental questions mentioned earlier. Such reconstruction program was initiated~\cite{Rovelli96} and had some successes, for example, in deriving the Schrodinger Equation.  Inspired by the RQM principles, an alternative reformulation of quantum mechanics was proposed~\cite{Yang2017, Yang2017_2}. The reformulation is based on two basic ideas. 1.)Relational properties between the two quantum systems are the most fundamental elements to formulate quantum mechanics. 2.)A physical measurement of a quantum system is a probe-response interaction process. Thus, the framework to calculate the probability of an outcome when measuring a quantum system should model this bidirectional process. This implies the probability can be derived from product of two quantities with each quantity associated with a unidirectional process. Such quantity is defined as relational probability amplitude. Specifically, the probability of a measurement outcome is proportional to the summation of probability amplitude product from all alternative measurement configurations. The properties of quantum systems, such as superposition and entanglement, are manifested through the rules of counting the alternatives. As results, traditional quantum mechanics formulations can be rediscovered but with new insights on the origin of quantum probability. Schrodinger Equation is recovered when there is no entanglement in the relational probability amplitude matrix~\cite{Yang2017}. On the other hand, when there is change in the entanglement measure, the quantum measurement theory is obtained~\cite{Yang2017_2}. In essence, quantum mechanics demands a new set of rules to calculate measurement probability from an interaction process. }

\replaced{What is missing in our initial reformulation~\cite{Yang2017} is an explicit calculation of the relational probability amplitude, which is at the heart of the framework. In this paper, we choose path integral method to calculate the relational probability amplitude for the following motivations. First, path integral offers an intuitive physical picture to calculate abstract quantity such as probability amplitude by summation over alternative trajectory paths. Since both path integral and our formulation are based on an idea of summation over alternatives, the technique in path integral can be naturally borrowed here. Second, path integral is a well-developed theory that had been successfully applied in other physical theories such as quantum field theory. By connecting the relational formulation of quantum mechanics to path integral, we wish to extend the relational formulation to more advance quantum theory in the future. Third, we can bring back new insight to the path integral itself by explaining why measurement probability can be calculated as modulus square of the probability amplitude. The outcome of implementing our formulation using path integral are fruitful, as shown in later sections.}{Although the concept of relational probability amplitude is useful to derive the quantum probability, its physical meaning is not obvious to understand. It is desirable to find an explicit calculation of the relational probability amplitude. It turns out that the path integral method can be used to achieve this goal and is briefly described in Ref.~\cite{Yang2017}. In this paper, the significance of the path integral implementation of relational quantum mechanics is fully developed.} Besides providing the physical meaning of relation probability amplitude, the path integral formulation also has interesting applications. For instance, it can describe the history of a quantum system that has interacted with a series of measuring systems in sequence. As a result, the double slit experiment can be elegantly explained from the formulation developed here. More significantly, the coordinator representation of the reduced density matrix derived from this implementation allows us to develop a method to calculate entanglement entropy using path integral approach. We propose a criterion on whether there is entanglement between the system and external environment based on the influence functional. This enables us to calculate entanglement entropy of a physical system that interacts with classical fields, such as an electron in an electromagnetic field. \deleted{Since entanglement entropy is an important concept in quantum information theory, the method described here may lead to new insight on the information aspect of a quantum system that interacts with classical fields.} 

The paper is organized as following. We first review the relational formulation of quantum mechanics in Section \ref{sec:RQM} \added{with a clearer explanation of the framework to calculate the measurement probability}. In Section \ref{sec:PI} the path integral implementation of the relational probability amplitude is presented. \deleted{It is shown to be compatible with the traditional path integral quantum mechanics, particularly on the definition of the influence functional.} Section \ref{sec:history} generalizes the formulation to describe the history of quantum state for the observed system that has interacted with a series of measuring systems in sequence. \added{The calculation confirms the idea that a quantum state encodes information from early measurements.}. The formulation is applied to explain the double slit experiment in Section \ref{sec:doubleslit}. Section \ref{sec:entropy} introduces a method to calculate entanglement entropy between the interacting systems based on the path integral implementation. A criterion to determine whether entanglement entropy vanishes based on properties of the influence functional is discussed in Section \ref{sec:conclusion}. Section \ref{sec:conclusion} \added{explores the idea of combining the present formulations with QRF theory, and} summarizes the conclusions. 

\section{Relational Formulation of Quantum Mechanics}
\label{sec:RQM}

\subsection{Terminologies}
\label{subsec:definition}
A \textit{Quantum System}, denoted by symbol $S$, is an object under study and follows the laws of quantum mechanics. An \textit{Apparatus}, denoted as $A$, can refer to the measuring devices, the environment that $S$ is interacting with, or the system from which $S$ is created. All systems are quantum systems, including any apparatus. Depending on the selection of observer, the boundary between a system and an apparatus can change. For example, in a measurement setup, the measuring system is an apparatus $A$, the measured system is $S$. However, the composite system $S+A$ as a whole can be considered a single system, relative to another apparatus $A'$. In an ideal measurement to measure an observable of $S$, the apparatus is designed in such a way that at the end of the measurement, the pointer state of $A$ has a distinguishable, one to one correlation with the eigenvalue of the observable of $S$.

The definition of \textit{Observer} is associated with an apparatus. An observer, denoted as $\cal{O}$, is an entity who can operate and read the pointer variable of the apparatus. Whether or not this observer is a quantum system is irrelevant in our formulation. An observer is defined to be physically local to the apparatus he associates with. This prevents the situation that $\cal{O}$ can instantaneously read the pointer variable of the apparatus that is space-like separated from $\cal{O}$. 

In the traditional theory proposed by von Neumann~\cite{Neumann} \added{for an ideal quantum measurement}, both the measured system $S$ and the measuring apparatus $A$ follow the same quantum mechanics laws. During the measurement process, $A$ probes (or, disturbs) $S$. Such interaction alters the state of $S$, which in turn responses to $A$ and alters the state of $A$ as well. As a result, a correlation is established between $S$ and $A$, allowing the measurement result for $S$ to be inferred from the pointer variable of $A$. 

\added{A \textit{Quantum Reference Frame (QRF)} is a quantum system where all the descriptions of the relational properties between $S$ and $A$ is referred to. There can be multiple QRFs. How the descriptions are transformed when switching QRFs is not in the scope of this study. But we expect the theories developed in Ref.~\cite{Brukner, Hoehn2018, Yang2020} can be applicable here. In this paper, we only consider the description relative to one QRF, denoted as $F$. It is also possible to choose $A$ as the reference frame. In that case, $F$ and $A$ are the same quantum system in a measurement~\cite{Brukner, Yang2020}. Fig. 1 shows a schematic illustration of the entities in the relational formulation.}

\begin{figure}
\begin{center}
\includegraphics[scale=2.2]{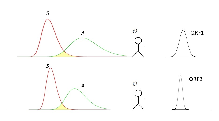}
\caption{\added{Schematic illustration of the entities for the terminologies. The overlapping of the measured system $S$ and apparatus $A$ represents there is interaction in a measurement. The relational properties between $S$ and $A$ must be described relative a QRF. There can be multiple QRFs. Selecting a different QRF, $\cal{O}$ can have a different description of the relational properties in a quantum measurement event.}}
\label{fig:1}       
\end{center}
\end{figure}

A \textit{Quantum State} of $S$ describes the complete information an observer $\cal{O}$ can know about $S$. A quantum state encodes the relational information of $S$ relative to $A$ or other systems that $S$ previously interacted with~\cite{Rovelli07}. The information encoded in the quantum state is the complete knowledge an observer can say about $S$, as it determines the possible outcomes of next measurement. As we will explain later, the state for a quantum system is not a fundamental concept. Instead, it is a derivative concept from the relational properties. When next measurement with another apparatus $A'$ is completed, the description of quantum state is updated to be relative to $A'$.

\subsection {\replaced{Measurement Probability}{Basic Formulation}}
\label{sub:measProb}
\added{In essence, quantum mechanics is a theory to predict the probability of future potential measurement given a prepared quantum system.} The relational formulation of quantum mechanics~\cite{Yang2017} is based on a detailed analysis of the interaction process during measurement of a quantum system. \added{To begin with, we assume a QRF, $F$, is chosen to describe the quantum measurement event.} From experimental observations, we know that a measurement of a variable on a quantum system yields multiple possible outcomes randomly. Each potential outcome is obtained with a certain probability. We call each measurement with a distinct outcome a quantum measurement event. Denote these alternatives potential outcome with a set of kets $\{|s_i\rangle\}$ for $S$, where ($i=0,\ldots ,N-1)$, and a set of kets $\{|a_j\rangle\}$ corresponding to the pointer readings for $A$, where ($j=0,\ldots ,M-1)$. A potential measurement outcome is represented by a pair of kets $(|s_i\rangle, |a_j\rangle)$. \deleted{Second, a physical measurement is a bidirectional process, the measuring system and the measured system interact and modify the state of each other.}  The probability of finding a potential measurement outcome represented by a pair of kets $(|s_i\rangle, |a_j\rangle)$, is denoted as $p_{ij}$. \added{The framework to calculate the measurement probability $p_{ij}$ is the central point of the formulation in Ref.~\cite{Yang2017}. Here we will give a more detailed explanation of this framework in a non-selective measurement setting. A physical measurement is a bidirectional process. In such process, the measuring system probes the measured system, and the measured system responses back to the measuring system. Through such interaction the states of both systems are modified and the correlation information is encoded. Although the bidirectional interaction process is well known, the following realization is not fully appreciated in the research literature.
\begin{displayquote}
\textit{Because a quantum measurement is a bidirectional process, the calculation of the probability of a measurement outcome must faithfully model such bidirectional process.}
\end{displayquote}

The bidirectional process does not necessarily imply two sequential steps. Instead, the probing and responding processes are understood as two aspects of a complete process in a measurement event. We can use a classical probability problem to analogize this. Suppose tossing a special coin gets a face up with probability of $p$. Let us consider a measurement process that requires tossing two such coins in the same time, and the measurement is successful if one coin facing up and one coin facing down. We ask what is the probability of a successful measurement event. The answer is to multiple two probability quantities together, $p(1-p)$. In the similar manner, given the bidirectional process in a quantum measurement event, the observable measurement probability should be calculated as a product of two quantities of weights. One weight quantity is associated with the probing process from $A\to S$, denoted as $Q^{A\rightarrow S} (|a_j\rangle \cap |s_i\rangle)$, and the other is associated with the responding process from $S\to A$, denoted as $R^{S\rightarrow A} (|s_i\rangle \cap |a_j\rangle)$, so that
\begin{equation}
\label{prob1}
    p_{ij} \propto Q^{A\rightarrow S} (|a_j\rangle \cap |s_i\rangle)R^{S\rightarrow A} (|s_i\rangle \cap |a_j\rangle)
\end{equation}
The difference between analogy of the classical coin tossing measurement and a quantum measurement is that in a quantum measurement, each aspect of the process itself is not necessarily assigned a non-negative real number. To see this, let us analyze the factors that the weight $Q^{A\rightarrow S} (|a_j\rangle \cap |s_i\rangle)$ depends on. Intuitively, this quantity depends on three factors:}
\begin{enumerate}
\item \added{Likelihood of finding system S in state $|s_i\rangle$ without interaction;}
\item \added{Likelihood of finding system A in state $|a_j\rangle$ without interaction;}
\item \added{A factor that alters the above two likelihoods due to the passing of physical elements such as energy and momentum from $A\to S$ in the probing process.}
\end{enumerate}
\added{
Similarly, the other weight quantity $R^{S\rightarrow A} (|s_i\rangle \cap |a_j\rangle)$ depends on the similar first two factors, and the third factor that is due to the passing of physical elements from $S\to A$ in the responding process. The probing and responding processes are two distinct aspects that constitute a measurement interaction. However, only the measurement probability is observable, the unidirectional process of probing or responding alone is not observable. There is no reason to require each factor must be a non-negative real value, as long as the final calculation result of $p_{ij}$ is still a non-negative real number. The requirement of a quantity being a non-negative real value is only applicable when the quantity is a measurable one, such as $p_{ij}$. We summarize this crucial but subtle point as following:
\begin{displayquote}
\textit{Measurement probability of an observable event must be a non-negative real number. However, the requirement of being a non-negative real number is not applicable to non-measurable quantities for sub-processes that constitute a quantum measurement.}
\end{displayquote}
It is worth to mention that this approach to construct the measurement probability is an operational one based on the probing and responding model. In theory, there can be different models that yield the same probability value.}

Next, we require that the probability $p_{ij}$ should be symmetric with respect to either $S$ or $A$. What this means is that the probability is the same for both processes $|a_j\rangle \rightarrow |s_i\rangle \rightarrow |a_j\rangle$ that is viewed from $A$ and $|s_i\rangle \rightarrow |a_j\rangle \rightarrow |s_i\rangle$ that is viewed from $S$. A natural way to meet this requirement is to represent the two weight quantities as matrix elements, i.e., $Q^{A\rightarrow S} (|a_j\rangle \cap |s_i\rangle) = Q^{AS}_{ji}$, and $R^{S\rightarrow A} (|s_i\rangle \cap |a_j\rangle) = R^{SA}_{ij}$. This implies $p_{ij}$ can be expressed as product of two numbers,
\begin{equation}
\label{probprod}
    p_{ij} \propto Q_{ji}^{AS}R_{ij}^{SA}. 
\end{equation}
As just discussed, $Q_{ji}^{AS}$ and $R_{ij}^{SA}$ are not necessarily real non-negative number since each number alone only models a unidirectional process which is not a complete measurement process thus non-observable. On the other hand, $p_{ij}$ must be a non-negative real number since it models an observable measurement process. To satisfy such requirement, we further assume
\begin{equation}
\label{conjugate}
    Q_{ji}^{AS} = (R_{ij}^{SA})^*.
\end{equation}
Written in a different format, $Q_{ji}^{AS} = (R^{SA})^\dag_{ji}$. This means $Q^{AS} = (R^{SA})^\dag$. \added{Eq.(\ref{conjugate}) can be justified as following. The three dependent factors for $Q_{ji}^{AS}$ and $R_{ij}^{SA}$ are related to each other respectively. The likelihood of finding system $S$ in state $|s_i\rangle$ and system $A$ in state $|a_j\rangle$ without interaction are the same. The third factor is triggered by passing physical elements during interaction. There are conservation laws such as energy conservation and momentum conservation during interaction. Conceivably, the third factors for $Q_{ji}^{AS}$ and $R_{ij}^{SA}$ must be equal in absolute value, but may be different in phase. With all these considerations, it is reasonable to assume $|Q_{ji}^{AS}| = |R_{ij}^{SA}|$. We choose $Q_{ji}^{AS} = (R_{ij}^{SA})^*$ so that $p_{ij}=Q_{ji}^{AS}R_{ij}^{SA}$ is a non-negative real number. These justifications will be clearer later in the path integral implementation.} With (\ref{conjugate}), Eq.(\ref{probprod}) then becomes 
\begin{equation}
    p_{ij} = |R^{SA}_{ij}|^2/\Omega
\end{equation}
where $\Omega$ is a real number normalization factor. $Q_{ji}^{AS}$ and $R_{ij}^{SA}$ are called \textit{relational probability amplitudes}. Given the relation in Eq.(\ref{conjugate}), we will not distinguish the notation $R$ versus $Q$, and only use $R$. \added{Note that $Q_{ji}^{AS}$ and $R_{ij}^{SA}$ are relative to the QRF $F$. A more accurate notation needs to explicitly call out the dependency on $F$. This is omitted here since we are not studying how $Q_{ji}^{AS}$ and $R_{ij}^{SA}$ are transformed when switching QRF from $F$ to a different QRF in the present works. This notation of not calling out $F$ explicitly will remain throughout the rest of this paper.

To manifest the unique bidirectional process of quantum measurement, let us compare it with a measurement in classical mechanics. In classical mechanics, for a fixed measurement setting and exact same preparation of the measured system, we expect the measurement result to be deterministic and with only one outcome. In the mathematical notation, this means $Q_{00}^{AS}=R_{00}^{SA}=1$. Furthermore, a classical measurement process is not considered to alter the state of the measured system. Rather, the measured system $S$ alters the pointer variable in the measurement apparatus $A$. In a sense this is just a unidirectional process. Therefore, the dependency on the bidirectional process to compute the measurement probability is unique to quantum mechanics. This unique process is not well analyzed in the research literature except the transaction interpretation of quantum mechanics that shares the idea that a quantum event is a bidirectional transaction~\cite{Cramer}. However, there are fundamental differences between the transaction model and the measurement probability framework presented here, which will be discussed in details in section \ref{subsec:transaction}. }

\subsection {\added{Wave Function and Density Matrix}}
\label{sec:wave}
The relational matrix $R^{SA}$ gives the complete description of $S$. It provides a framework to derive the probability of future measurement outcome. Although $R^{SA}_{ij}$ is a probability amplitude, not a probability real number, we assume it follows certain rules in the classical probability theory, such as multiplication rule, and sum of alternatives in the intermediate steps.

The set of kets $\{|s_i\rangle\}$, representing distinct measurement events for $S$, can be considered as \textit{eigenbasis} of Hilbert space ${\cal H}_S$ with dimension $N$, and $|s_i\rangle$ is an eigenvector. Since each measurement outcome is distinguishable, $\langle s_i|s_j\rangle = \delta_{ij}$. Similarly, the set of kets $\{|a_j\rangle\}$ is eigenbasis of Hilbert space ${\cal H}_A$ with dimension $N$ for the apparatus system $A$. The bidirectional process $|a_j\rangle \rightleftharpoons |s_i\rangle$ is called a \textit{potential measurement configuration.} \replaced{A potential measurement configuration comprises possible eigen-vectors of $S$ and $A$ that involve in the measurement event, and the relational weight quantities. It can be represented by $\Gamma_{ij}: \{|s_i\rangle, |a_j\rangle, R^{SA}_{ij}, Q^{AS}_{ji}\}$.}{in the joint Hilbert space ${\cal H}_S \oplus {\cal H}_A$} 

To derive the properties of $S$ based on the relational $R$, we examine how the probability of measuring $S$ with a particular outcome of variable q is calculated. It turns out such probability is proportional to the sum of weights from all applicable measurement configurations, where the weight is defined as the product of two relational probability amplitudes corresponding to the applicable measurement configuration. Identifying the applicable measurement configuration manifests the properties of a quantum system. For instance, before measurement is actually performed, we do not know that which event will occur to the quantum system since it is completely probabilistic. It is legitimate to generalize the potential measurement configuration as $|a_j\rangle \rightarrow |s_i\rangle \rightarrow |a_k\rangle$. In other words, the potential measurement configuration \deleted{in the joint Hilbert space} starts from $|a_j\rangle$, but can end at $|a_j\rangle$, or any other event, $|a_k\rangle$. Indeed, the most general form of potential measurement configuration in a bipartite system can be $|a_j\rangle \rightarrow |s_m\rangle \rightarrow |s_n\rangle \rightarrow |a_k\rangle$. Correspondingly, we generalize Eq.(\ref{probprod}) by introducing a quantity for such configuration,
\begin{equation}
\label{micAction}
W_{jmnk}^{ASSA} = Q^{AS}_{jm}R^{SA}_{nk} = (R^{SA}_{mj})^*R^{SA}_{nk}.
\end{equation}
The second step utilizes Eq.(\ref{conjugate}). This quantity is interpreted as a weight associated with the potential measurement configuration $|a_j\rangle \rightarrow |s_m\rangle \rightarrow |s_n\rangle \rightarrow |a_k\rangle$. Suppose we do not perform actual measurement and inference information is not available, the probability of finding $S$ in a future measurement outcome can be calculated by summing $W_{jmnk}^{ASSA}$ from all applicable alternatives of potential measurement configurations. 

With this framework, the remaining task to calculate the probability is to correctly count the \textit{applicable} alternatives of potential measurement configuration. This task depends on the expected measurement outcome. For instance, suppose the expected outcome of an ideal measurement is event $|s_i\rangle$, i.e., measuring variable $q$ gives eigenvalue $q_i$. The probability of event $|s_i\rangle$ occurs, $p_i$, is proportional to the summation of $W_{jmnk}^{ASSA}$ from all the possible configurations related to $|s_i\rangle$. Mathematically, we select all $W_{jmnk}^{ASSA}$ with $m=n=i$, sum over index $j$ and $k$, and obtain the probability $p_i$.
\begin{equation}
\label{probability}
p_i \propto \sum_{j,k=0}^M (R^{SA}_{ij})^*R^{SA}_{ik}= |\sum_{j} R^{SA}_{ij}|^2.
\end{equation}
This leads to the definition of wave function $\varphi_i = \sum_{j} R_{ij}$, so that $p_i=|\varphi_i|^2$. The quantum state of $S$ can be described either by the relational matrix $R$, or by a set of variables $\{\varphi_i\}$. The vector state of $S$ relative to $A$, is $|\psi\rangle_S^A = (\varphi_0, \varphi_1,\ldots, \varphi_N)^T$ where superscript $T$ is the transposition symbol. More specifically, \added{we can define}
\begin{equation}
\label{WF}
|\psi\rangle_S^A = \sum_i\varphi_i |s_i\rangle  \quad \textrm{where } \varphi_i = \sum_{j} R_{ij}.
\end{equation}
The justification for the above definition is that the probability of finding $S$ in eigenvector $|s_i\rangle$ in future measurement can be calculated from it by defining a projection operator $\hat{P}_i=|s_i\rangle\langle s_i|$. Noted that $\{|s_i\rangle\}$ are orthogonal eigenbasis, the probability is rewritten as:
\begin{equation}
\label{probability3}
p_i=\langle\psi|\hat{P}_i|\psi\rangle = |\varphi_i|^2
\end{equation}

Eqs.(\ref{probability}) and (\ref{WF}) \replaced{show why the measurement probability equals the modulus square of wave function. They are derived based on two conditions. First, the probability is calculated by modeling the bidirectional measurement process; Second, there are indistinguishable alternatives of potential measurement configurations in computing the probability. When the alternatives are distinguishable, even through inference from correlation information with another quantum system, }{introduced on the condition that there is no entanglement (See Section \ref{subsec:entanglement} for the definition of entanglement) between quantum system $S$ and $A$. If there is entanglement between them,} the summation in Eq.(\ref{probability}) over-counts the applicable alternatives of measurement configurations and should be modified accordingly. A more generic approach to describe the quantum state of $S$ is the reduced density matrix formulation, which is defined as
\begin{equation}
\label{reducedRho}
\rho_S=RR^\dag
\end{equation}
The probability $p_i$ is calculated using the projection operator $\hat{P}_i=|s_i\rangle\langle s_i|$
\begin{equation}
\label{indirectProb}
p_i = Tr_S(\hat{P}_i\hat{\rho}_S) = \sum_j|R_{ij}|^2.
\end{equation}

The effect of a quantum operation on the relational probability amplitude matrix can be expressed through an operator. \replaced{By defining}{Defined} an operator $\hat{M}$ in Hilbert space ${\cal H}_S$ as $M_{ij} = \langle s_i|\Hat{M}|s_k\rangle$, \replaced{one can obtain the new relational probability amplitude matrix $R^{SA}_{new}$ from the transformation of the initial matrix $R^{SA}_{init}$}{The new relational probability amplitude matrix is obtained by}
\begin{equation}
    \label{operator1}
    \begin{split}
    (R^{SA}_{new})_{ij} &= \sum_{k}M_{ik}(R^{SA}_{init})_{kj}, \quad \text{or} \\ R_{new} &= MR_{init}.
    \end{split}
\end{equation}
Consequently, the reduced density becomes
\begin{equation}
    \label{operator3}
    \rho_{new} = R_{new}(R_{new})^\dag = M\rho_{init}M^\dag.
\end{equation}

\subsection{Entanglement Measure}
\label{subsec:entanglement}
The description of $S$ using the reduced density matrix $\rho_S$ is valid regardless there is entanglement between $S$ and $A$. To determine whether there is entanglement between $S$ and $A$, a parameter to characterize the entanglement measure should be introduced. There are many forms of entanglement measure~\cite{Nelson00, Horodecki}, the simplest one is the von Neumann entropy, which is defined as
\begin{equation}
    \label{vonNeumann2}
    H(\rho_S)  = -Tr(\rho_S ln(\rho_S)).
\end{equation}
Denote the eigenvalues of the reduced density matrix $\rho_S$ as $\{\lambda_i\}, i=0,\ldots, N$, the von Neumann entropy is calculated as
\begin{equation}
    \label{vonNeumann}
    H(\rho_S)  = -\sum_i\lambda_i ln\lambda_i.
\end{equation}
A change in $H(\rho_S)$ implies there is change of entanglement between $S$ and $A$. Unless explicitly pointed out, we only consider the situation that $S$ is described by a single relational matrix $R$. In this case, the entanglement measure $E=H(\rho_S)$.

$H(\rho_S)$ enables us to distinguish different quantum dynamics. Given a quantum system $S$ and an apparatus $A$, there are two types of the dynamics between them. In the first type of dynamics, there is no physical interaction and no change in the entanglement measure between $S$ and $A$. $S$ is not necessarily isolated in the sense that it can still be entangled with $A$, but the entanglement measure remains unchanged. This type of dynamics is defined as \textit{time evolution}. In the second type of dynamics, there is a physical interaction and correlation information exchange between $S$ and $A$, i.e., the von Neumann entropy $H(\rho_S)$ changes. This type of dynamics is defined as \textit{quantum operation}. \textit{Quantum measurement} is a special type of quantum operation with a particular outcome. Whether the entanglement measure changes distinguishes a dynamic as either a time evolution or a quantum operation~\cite{Yang2017, Yang2017_2}. 

\section{Results}
\subsection{\added{Motivation for Path Integral Implementation}}
\label{sec:motivation}
\added{
As shown in the previous sections, the relational probability amplitude $R_{ij}$ provides a complete description of the quantum system relative to a reference system. It is natural to ask what the physical meaning of this quantity is and how to mathematically calculate it. A concrete implementation of the relational quantum mechanics depends on how $R_{ij}$ is calculated.

In the present works, path integral is chosen to be the implementation method for several reasons. First, traditional path integral offers a physical model to compute the probability amplitude through the summation of contributions from all alternative paths. A physical picture can help to clarify the abstract concepts. Second, mathematically, the technique of summation over alternatives in path integral is similar to the method in calculating the measurement probability, as shown in Eq.(\ref{probability}). The mathematical method in path integral can be borrowed intuitively to calculate the relational probability amplitude. Third, path integral is a well-developed theory that had been successfully applied in other physical theories such as quantum field theory. By connecting the relational formulation of quantum mechanics to path integral, we wish to extend the relational formulation to more advance quantum theory in the future.

It is important to point out that we are going to only borrow the idea from path integral on how the relation probability amplitude is calculated, nothing else. Because once the relational probability amplitude is calculated, the basic quantum theory such as the Schr\"{o}dinger equation, measurement formulation, can be recovered as shown in Refs.~\cite{Yang2017, Yang2017_2}. In addition, the path integral implementation presented here brings back valuable inside to the path integral formulation of quantum mechanics itself. In Feynman's original paper~\cite{Feynman48}, the fact that measurement probability equals to the modulus square of the probability amplitude, $p_i=|\phi|^2$, is introduced as a postulate. The justification is that whenever there are indistinguishable alternatives during a measurement event, such postulate holds~\cite{Feynman05}. However, as we mentioned in Section \ref{sec:wave}, $p_i=|\phi|^2$ is derived based on two conditions. Indistinguishable alternatives is just one of them. Modeling the bidirectional measurement process in calculating the probability is the other key reason. In a sense, Feynman's explanation on $p_i=|\phi|^2$ is incomplete. By implementing the relation probability
amplitude using path integral formulation, we help to complete the justification. When the alternatives can be indistinguishable, they are called ``interfering alternatives". When the alternatives are distinguishable such as in the case of entanglement between $S$ and $A$, they are called ``exclusive alternatives". Supposed each alternative is assigned a weight, the rules to calculate the probability can be summarized as
\begin{displayquote}
\textbf{Probability for Alternatives} \textit{To calculate the probability for interfering alternatives, one first takes the summation of weight for each alternative, then takes the modulus square of the summation. To calculate the probability for exclusive alternatives, one first takes the modulus square of weight for each alternative, then takes the summation.}
\end{displayquote}
In the other words, the order of taking modulus square and taking summation is swapped for both cases, as clearly shown in (\ref{probability}) and (\ref{indirectProb}). We will see in Section \ref{sec:doubleslit} that this rule is further manifested when explaining the double slit experiment.
}

\subsection{Path Integral Implementation}
\label{sec:PI}

\deleted{This section shows that the relational probability amplitude can be explicitly calculated using the Path Integral formulation. }Without loss of generality, the following discussion just focuses on one dimensional space-time quantum system. In the Path Integral formulation, the probability to find a quantum system moving from a point $x_a$ at time $t_a$ to a point $x_b$ at time $t_b$ is postulated to be the absolute square of a probability amplitude, i.e., $P(b, a)=|K(b, a)|^2$ (as mentioned earlier, such postulate is not needed in our reformulation). The probability amplitude is further postulated as the sum of the contributions of phase from all alternative paths~\cite{Feynman05}:
\begin{equation}
\label{PIWF2}
K(b, a)=\int_{a}^{b}e^{(i/\hbar)S_p(x(t))}{\cal{D}}x(t)
\end{equation}
where ${\cal{D}}x(t)$ denotes integral over all possible paths from point $x_a$ to point $x_b$. It is the wave function for $S$ moving from $x_a$ to $x_b$~\cite{Feynman05}. The wave function of finding the particle in a region ${\cal{R}}_b$ previous to $t_b$ can be expressed as
\begin{equation}
    \label{PIWF3}
    \varphi(x_b)=\int_{{\cal{R}}_b}e^{(i/\hbar)S_p(x(t))}{\cal{D}}x(t)
\end{equation}
where $x_b$ is the position of particle at time $t_b$. The integral over region ${\cal{R}}_b$ can be interpreted as integral of all paths ending at position $x_b$, with the condition that each path lies in region ${\cal{R}}_b$ which is previous to time $t_b$. The rational of such definition can be found in Feynman's original paper~\cite{Feynman48}.

Now let's consider how the relational matrix element can be formulated using similar formulation. Define ${\cal{R}}_b^S$ is the region of finding system $S$ previous to time $t_b$, and ${\cal{R}}_b^A$ is the region of finding measuring system $A$ previous to time $t_b$. We denote the matrix element as $R(x_b; y_b)$, where the coordinates $x_b$ and $y_b$ act as indices to the system $S$ and apparatus $A$, respectively. Borrowing the ideas described in Eq.(\ref{PIWF2}), we propose that the relational matrix element is calculated as
\begin{equation}
    \label{PIR}
    \begin{split}
    R(x_b, y_b) &= \int_{{\cal{R}}_b^S} \int_{{\cal{R}}_b^A}e^{(i/\hbar)S^{SA}_p(x(t), y(t))} {\cal{D}}x(t){\cal{D}}y(t)
    \end{split}
\end{equation}
where the action $S_p^{SA}(x(t), y(t))$ consists three terms
\begin{equation}
    \label{actions}
    \begin{split}
    S^{SA}_p(x(t), y(t)) &= S^S_p(x(t)) + S^A_p(y(t)) \\
    & + S^{SA}_{int}(x(t), y(t)).
    \end{split}
\end{equation}
The last term is the action due to the interaction between $S$ and $A$ when each system moves along its particular path. \added{The phase contributions from each of the action terms are corresponding to the three factors mentioned in section \ref{sub:measProb}. For instance, $e^{(i/\hbar)S^{S}_p}$ is corresponding to the contribution to the likelihood of finding $S$ in position $x_a$, and $e^{(i/\hbar)S^{SA}_{int}}$ is corresponding to the factor due to responding from $S\to A$. We would like to remind that the expressions in (\ref{PIR}) and (\ref{actions}) are relative to the QRF $F$ but we do not explicitly call out this dependency without loss of clarity. }

We can validate (\ref{PIR}) by deriving formulation that is consistent with traditional path integral. Suppose there is no interaction between $S$ and $A$. The third term in Eq.(\ref{actions}) vanishes. (\ref{PIR}) is decomposed to product of two independent terms,
\begin{equation}
    \label{PIR2}
    \begin{split}
    R(x_b, y_b) & = \int_{{\cal{R}}_b^S} e^{(i/\hbar)S^{S}_p(x(t)) }{\cal{D}}x(t)\\
    & \times \int_{{\cal{R}}_b^A} e^{(i/\hbar)S^{A}_p(y(t)) }{\cal{D}}y(t)
    \end{split}
\end{equation}
Noticed that the coordinates $y_a$ and $y_b$ are equivalent of the index $j$ in Eq.(\ref{WF}), the wave function of $S$ can be obtained by integrating $y_b$ over Eq.(\ref{PIR2})
\begin{equation}
    \label{PIWF4}
    \begin{split}
    \varphi(x_b) & = \int^{\infty}_{-\infty}R(x_b, y_b)dy_b \\
    &=\int_{{\cal{R}}_b^S} e^{(i/\hbar)S^{S}_p(x(t)) }{\cal{D}}x(t)\\
    &\times \int^{\infty}_{-\infty}\int_{{\cal{R}}_b^A} e^{(i/\hbar)S^{A}_p(y(t))}{\cal{D}}y(t)dy_b \\
    &=c\int_{{\cal{R}}_b^S} e^{(i/\hbar)S^{S}_p(x(t))}{\cal{D}}x(t)
    \end{split}
\end{equation}
where constant $c$ is the integration result of the second term in step two. The result is the same as Eq.(\ref{PIWF3}) except an unimportant constant.

Next, we consider the situation that there is entanglement between $S$ and $A$ as a result of interaction. The third term in Eq.(\ref{actions}) does not vanish. We can no longer define a wave function for $S$. Instead, a reduced density matrix should be used to describe the state of the particle, $\rho = RR^\dag$. Similar to (\ref{PIR}), the element of the reduced density matrix is 
\begin{equation}
    \label{PIRho}
    \begin{split}
    \rho(x_b, x'_b) &= \int dy_b\int_{{\cal{R}}_b^S}\int_{{\cal{R}}_{b'}^S}\int_{{\cal{R}}_b^A}\int_{{\cal{R}}_b^A}e^{(i/\hbar)\Delta S} \\
    &\times {\cal{D}}x(t){\cal{D}}x'(t){\cal{D}}y(t){\cal{D}}y'(t)\\
    \text{where}\quad \Delta S &= S^S_p(x(t)) - S^S_p(x'(t)) \\ 
    & + S^A_p(y(t))  - S^A_p(y'(t)) \\
    & + S^{SA}_{int}(x(t), y(t)) \\
    & - S^{SA}_{int}(x'(t), y'(t)).
    \end{split}
\end{equation}
Here $x_b=x(t_b)$ and $x'_b=x'(t_b)$. The path integral over ${\cal{D}}y'(t)$ takes the same region  (therefore same end point $y_b$) as the path integral over ${\cal{D}}y(t)$. After the path integral, an integral over $y_b$ is performed. Eq.(\ref{PIRho}) is equivalent to the $J$ function introduced in Ref~\cite{Feynman05}. We can rewrite the expression of $\rho$ using the \textit{influence functional}, $F(x(t), x'(t))$,
\begin{equation}
    \label{PIIF}
    \begin{split}
    \varrho(x_b, x'_b) &= \frac{1}{Z} \int_{{\cal{R}}_b^S}\int_{{\cal{R}}_{b'}^S}e^{(i/\hbar)[S^S_p(x(t)) - S^S_p(x'(t))]} \\
    &\times F(x(t), x'(t)){\cal{D}}x(t){\cal{D}}x'(t) \\
    F(x(t), x'(t)) & = \int dy_b\int\int_{{\cal{R}}_b^A}e^{(i/\hbar)\Delta S'}{\cal{D}}y(t){\cal{D}}y'(t)\\
    \text{where}\quad \Delta S' &= S^A_p(y(t))  - S^A_p(y'(t)) \\
    & + S^{SA}_{int}(x(t), y(t)) \\
    & - S^{SA}_{int}(x'(t), y'(t)).
    \end{split}
\end{equation}
where $Z=Tr(\rho)$ is a normalization factor to ensure $Tr(\varrho) = 1$. 

In summary, we show that the relational probability amplitude can be explicitly calculated through Eq.(\ref{PIR}). $R^{SA}_{ij}$ is defined as the sum of quantity $e^{iS_p/\hbar}$, where $S_p$ is the action of the composite system $S+A$ along a path. Physical interaction between $S$ and $A$ may cause change of $S_p$, which is the phase of the probability amplitude. But $e^{iS_p/\hbar}$ itself is a probabilistic quantity, instead of a quantity associated with a physical property. With this definition and the results in Section \ref{sec:RQM}, we obtain the formulations for wave function (\ref{PIWF4}) and the reduced density matrix (\ref{PIIF}) that are the consistent with those in traditional path integral formulation. Although the reduced density expression (\ref{PIRho}) is equivalent to the $J$ function in Ref~\cite{Feynman05}, it has richer physical meaning. For instance, we can calculate the entanglement entropy from the reduced density matrix. This will be discussed further in Section ~\ref{sec:entropy}.

\subsection{Interaction History of a Quantum System}
\label{sec:history}
One of the benefits of implementing the relational probability amplitude using path integral approach is that it is rather straightforward to describe the interaction history of a quantum system. Let's start with a simple case and later extend the formulation to a general case.

Suppose up to time $t_a$, a quantum system $S$ only interacts with a measuring system $A$. The detail of interaction is not important in this discussion. $S$ may interact with $A$ for a short period of time or may interact with $A$ for the whole time up to $t_a$. Assume that after $t_a$, there is no further interaction between $S$ and $A$. Instead $S$ starts to interact with another measuring system $A'$, up to time $t_b$. Denote the trajectories of $S, A, A'$ as $x(t), y(t), z(t)$, respectively. Up to time $t_a$, the relational matrix element is given by Eq.(\ref{PIR}),
\begin{equation}
    \label{PIR3}
    \begin{split}
    R(x_a, y_a) &= \int_{{\cal{R}}_a^S} \int_{{\cal{R}}_a^A}e^{\frac{i}{\hbar}S^{SA}_p(x(t), y(t))} {\cal{D}}x(t){\cal{D}}y(t).
    \end{split}
\end{equation}
Up to time $t_b$, \replaced{as $S$ is interacting with a different apparatus $A'$, the relational properties is described by a different relational matrix, with the matrix element as}{the relational matrix element becomes}
\begin{equation}
    \label{PIR4}
    \begin{split}
   & R(x_b, y_a, z_b) = \int_{{\cal{R}}_b^S} \int_{{\cal{R}}_a^A}\int_{{\cal{R}}_b^{A'}} {\cal{D}}x(t){\cal{D}}y(t){\cal{D}}z(t)\\
    & \times exp\{\frac{i}{\hbar}[S^{SA}_p(x(t), y(t))+S^{SA'}_p(x(t), z(t))]\} 
    \end{split}
\end{equation}
We can split region ${\cal{R}}_b^S$ into two regions, ${\cal{R}}_a^S$ and ${\cal{R}}_{ab}^S$, where ${\cal{R}}_{ab}^S$ is a region between time $t_a$ and time $t_b$. This split allows us to express $R(x_b, y_a, z_b)$ in terms of $R(x_a, y_a)$,
\begin{equation}
    \label{PIR5}
    R(x_b, y_a, z_b) =  \int R(x_a, y_a) K(x_a, x_b, z_b)dx_a,
\end{equation}
where 
\begin{equation}
    \label{KFunction}
    \begin{split}
    & K(x_a, x_b, z_b) \\
    & = \int_{{\cal{R}}_{ab}^S}\int_{{\cal{R}}_b^{A'}}{\cal{D}}x(t){\cal{D}}z(t)exp\{\frac{i}{\hbar}S^{SA'}_p(x(t), z(t))\} 
    \end{split}
\end{equation}
From Eq.(\ref{KFunction}), one can derive the reduced density matrix element for $S$,
\begin{equation}
    \label{PIRho2}
    \begin{split}
    \rho(x_b, x'_b)& =\int\int  R(x_b, y_a, z_b)R^*(x'_b, y_a, z_b)dy_adz_b\\
    & = \int\int\rho(x_a, x'_a) G(x_a, x'_a; x_b, x'_b) dx_a dx'_a
    \end{split}
\end{equation}
where
\begin{equation}
    \label{Gfunction}
    G(x_a, x_b; x'_a, x'_b) = \int K(x_a, x_b, z_b)K^*(x'_a, x'_b, z_b)dz_b.
\end{equation}
Normalizing the reduced density matrix element, we have
\begin{equation}
    \label{PIRho3}
    \varrho((x_b, x'_b)=\frac{1}{Z}\int\int\rho(x_a, x'_a) G(x_a, x'_a; x_b, x'_b) dx_a dx'_a
\end{equation}
where the normalization factor $Z=Tr(\rho)$.

The reduced density matrix allows us to calculate the probability of finding the system in a particular state. For instance, suppose we want to calculate the probability of finding the system in a state $\psi(x_b)$. Let $\Hat{P}=|\psi(x_b)\rangle\langle \psi(x_b)|$ the project operator for state $\psi(x_b)$. According to (\ref{indirectProb}), the probability is calculated as 
\begin{equation}
    \label{PIprob}
    \begin{split}
    p(\psi) &=Tr(\rho\Hat{P})= \langle \psi(x_b)|\varrho |\psi(x_b)\rangle\\
    &=\int\int \psi^*(x'_b)\psi(x_b)\varrho(x_b, x'_b) dx_bdx'_b
    \end{split}
\end{equation}
To find the particle at a particular position $\bar{x}_b$ at time $t_b$, we substitute $\psi(x_b)=\delta(x_b-\bar{x}_b)$ into Eq.(\ref{PIprob}),
\begin{equation}
    \label{PIdensityProb}
    \begin{split}
    p(\bar{x}_b) &= \int\int\varrho(x_b, x'_b)\delta (x_b - \bar{x}_b)\delta (x'_b - \bar{x}_b) dx_b dx'_b \\
    & = \varrho(\bar{x}_b, \bar{x}_b).
    \end{split}
\end{equation}

Suppose further that there is no interaction between $S$ and $A'$ after time $t_a$, the action $S_p^{SA'}(x(t), z(t))$ consists only two independent terms, $S^{SA'}_p(x(t), z(t)) = S^S_p(x(t)) + S^{A'}_p(z(t))$. This allows us to rewrite (\ref{KFunction}) as a product of two terms:
\begin{equation}
    \label{PIR7}
    \begin{split}
    & K(x_a, x_b, z_b) =  \int_{{\cal{R}}_{ab}^S}{\cal{D}}x(t)exp\{\frac{i}{\hbar}S^S_p(x(t))\} \\
    & \times  \int_{{\cal{R}}_b^{A'}}{\cal{D}}z(t)exp\{\frac{i}{\hbar}S^{A'}_p(z(t))\}\\
    & = K(x_a, x_b) \int_{{\cal{R}}_b^{A'}}{\cal{D}}z(t)exp\{\frac{i}{\hbar}S^{A'}_p(z(t))\}
    \end{split}
\end{equation}
Consequently, function $G(x_a, x'_a; x_b, x'_b)$ is rewritten as
\begin{equation}
    \label{PIR8}
    \begin{split}
    &G(x_a, x'_a; x_b, x'_b) = K(x_a, x_b)K^*(x'_a, x'_b) \\
    &\times \int dz_b\int\int_{{\cal{R}}_b^{A'}}{\cal{D}}z(t){\cal{D}}z'(t)exp\{\frac{i}{\hbar}[S^{A'}_p(z(t))-S^{A'}_p(z'(t))\}.
    \end{split}
\end{equation}
The integral in the above equation is simply a constant, denoted as $C$. Thus, $G(x_a, x'_a; x_b, x'_b) = CK(x_a, x_b)K^*(x'_a, x'_b)$. Substituting this into (\ref{PIRho3}), one obtains
\begin{equation}
    \label{PIRho4}
    \varrho(x_b, x'_b)=\frac{1}{Z}\int\int K(x_a, x_b)\rho(x_a, x_b)K^*(x'_a, x'_b)dx_a dx'_a.
\end{equation}
The constant $C$ is absorbed into the normalization factor. In next section, we will use (\ref{PIRho4}) and (\ref{PIdensityProb}) to explain the double slit experiment.

We wish to generalize (\ref{PIR4}) and (\ref{PIRho3}) to describe a series of interaction history of the quantum system $S$. Suppose quantum system $S$ has interacted with a series of measuring systems $A_1, A_2, \ldots, A_n$ along the history but interacts with one measuring system at a time. Specifically, $S$ interacts with $A_1$ up to time $t_1$. From $t_1$ to $t_2$, it only interacts with $A_2$. From $t_2$ to $t_3$, it only interacts with $A_3$, and so on. Furthermore, we assume there is no interaction among these measuring systems. They are all independent. Denote the trajectories of these measuring systems as $y^{(1)}(t), y^{(2)}(t), \ldots, y^{(n)}(t)$, and $y^{(1)}(t_1)=y^{(1)}_b, y^{(2)}(t_2)=y^{(2)}_b, \ldots, y^{(n)}(t_n)=y^{(n)}_b$. With this model, we can write down the relational matrix element for the whole history
\begin{equation}
    \label{history}
    \begin{split}
   & R^{(n)}(x_b^{(n)}, y^{(1)}_b, y^{(2)}_b, \ldots, y^{(n)}_b) \\
   & = \int_{{\cal{R}}_n^S}{\cal{D}}x(t) \prod_{i=1}^n\{\int_{{\cal{R}}_i^{A}}{\cal{D}}y^{(i)}(t)\}\\
    & \times exp\{\frac{i}{\hbar}\sum_{j=1}^n[S^{(j)}_p(x(t), y^{(j)}(t))]\} 
    \end{split}
\end{equation}
where ${\cal{R}}_n^S$ is the region of finding the measured system $S$ previous to time $t_n$, ${\cal{R}}_i^A$ is the region of finding measuring system $A_i$ between time $t_{i-1}$ to $t_i$. Action
\begin{equation}
    \label{action2}
    \begin{split}
    S^{(j)}_p(x(t), y^{(j)}(t))  &= \int_{t_{j-1}}^{t_j} L(x(t), \dot{x}(t), y^{(j)}(t), \dot{y}^{(j)}(t)) dt\\
    \end{split}
\end{equation}
is the integral of the Lagrangian over a particular path $p$ lying in region ${\cal{R}}_n^S\cup {\cal{R}}_j^A$. The reason the total action among $S$ and $A_1, A_2, \ldots, A_n$ is written as a summation of individual action between $S$ and $A_i$ is due to the key assumption that $S$ only interacts with one measuring system $A_i$ at a time, and the interaction with each of measuring system $A_i$ is independent from each other. This assumption further allows us to separate path integral over region ${\cal{R}}_n^S$ into two parts, one is integral over region previous to $t_{n-1}$, ${\cal{R}}_{n-1}^S$, and the other is the integral over region between $t_{n-1}$ and $t_{n}$, ${\cal{R}}_{n-1, n}^S$. Thus,
\begin{equation}
    \label{history2}
    \begin{split}
   & R^{(n)}(x_b^{(n)}, y^{(1)}_b, y^{(2)}_b, \ldots, y^{(n)}_b) \\
   & = \int R^{(n-1)}(x_b^{(n-1)}, y^{(1)}_b, y^{(2)}_b, \ldots, y^{(n-1)}_b) \\
   & \times K(x_b^{(n-1)}, x_b^{(n)},y^{(n)}_b) dx_b^{(n-1)}
    \end{split}
\end{equation}
and 
\begin{equation}
    \label{history3}
    \begin{split}
    K(x_b^{(n-1)}, x_b^{(n)},y^{(n)}_b) &= \int_{{\cal{R}}_{n-1,n}^S}{\cal{D}}x(t)\int_{{\cal{R}}_n^{A}}{\cal{D}}y^{(n)} \\
    & \times exp\{\frac{i}{\hbar}S^{(n)}_p(x(t), y^{(n)}(t))\} 
    \end{split}
\end{equation}
Similar to (\ref{PIRho}), the reduced density matrix element at $t_n$ is
\begin{equation}
    \label{PIRho5}
    \begin{split}
    \rho(x_b^{(n)}, x_{b'}^{(n)}) &=[\prod_{i=1}^{i=n}\int dy^{(i)}_b]  R^{(n)}(x_b^{(n)}, y^{(1)}_b, y^{(2)}_b, \ldots, y^{(n)}_b) \\
    & \times (R^{(n)}(x_b^{(n)}, y^{(1)}_b, y^{(2)}_b, \ldots, y^{(n)}_b))^*\\
    & = \int\int  \rho(x_b^{(n-1)}, x_{b'}^{(n-1)}) \\
    &\times G(x_b^{(n-1)}, x_b^{(n)}, x_{b'}^{(n-1)}, x_{b'}^{(n)}) dx_b^{(n-1)} dx_{b'}^{(n-1)} 
     \end{split}
\end{equation}
where
\begin{equation}
    \label{PIRho6}
    \begin{split}
    & G(x_b^{(n-1)}, x_b^{(n)}, x_{b'}^{(n-1)}, x_{b'}^{(n)}) = \\
    & \int K(x_b^{(n-1)}, x_b^{(n)},y^{(n)}_b)K^*(x_{b'}^{(n-1)}, x_{b'}^{(n)},y^{(n)}_b) dy^{(n)}_b .
    \end{split}
\end{equation}
The normalized version of the reduced density matrix element is given by 
\begin{equation}
    \varrho(x_b^{(n)}, x_{b'}^{(n)}) = \frac{\rho(x_b^{(n)}, x_{b'}^{(n)})}{Tr(\rho(t_n))}
\end{equation}
and the probability of finding $S$ in a position $\bar{x}_b^{(n)}$ at time $t_n$ is
\begin{equation}
    \label{PIdensityProb2}
    p(\bar{x}_b^{(n)})  = \varrho(\bar{x}_b^{(n)}, \bar{x}_b^{(n)}).
\end{equation}
Recall that $x_b^{(n)}, x_{b'}^{(n)}$ are two different positions of finding $S$ at time $t_n$. The probability of finding $S$ at position $\bar{x}_b^{(n)}$ is simply a diagonal element of the reduced density matrix.

\added{The formulation presented in this section confirms the idea that the relational probability amplitude matrix, and consequently the wave function of $S$, encodes the relational information along the history of interactions between $S$ and other systems $\{A_1, A_2,..., A_n\}$. The idea was original conceived in Ref.~\cite{Rovelli07}, but was not fully developed there.}

\subsection{The Double Slit Experiment}
\label{sec:doubleslit}
In the double slit experiment, an electron passes through a slit configuration at screen C and is detected at position $x$ of the destination screen B. Denote the probability that the particle is detected at position $x$ as $p_1=|\varphi_1(x)|^2$ when only slit $T_1$ is opened, and as $p_2=|\varphi_2(x)|^2$ when only slit $T_2$ is opened. If both slits are opened, the probability that the particle is detected at position $x$ after passing through the double slit is given by $p(x)=|\varphi_1(x)+\varphi_2(x)|^2$, which is different from $p_1+p_2 = |\varphi_1(x)|^2+|\varphi_2(x)|^2$. Furthermore, when a detector is introduced to detect exactly which slit the particle goes though, the probability becomes $|\varphi_1(x)|^2+|\varphi_2(x)|^2$. This observation was used by Feynman to introduce the concept of probability amplitude and the rule of calculating the measurement probability as the absolute square of the probability amplitude~\cite{Feynman05}. \replaced{The reason for this, according to Feynman, is that the alternatives of passing $T1$ or $T2$ is indistinguishable. Thus, these alternatives are ``interfering alternatives" instead of ``exclusive alternatives". However, how the probability $p(x)=|\varphi_1(x)+\varphi_2(x)|^2$ or $p(x)=|\varphi_1(x)|^2+|\varphi_2(x)|^2$ is calculated is not provided.}{However, why the probability is the absolute square of the probability amplitude is not explained in Ref.~\cite{Feynman05}. By recognizing the measurement probability is a directional process, the measurement probability is shown to be the absolute square of probability amplitude~\cite{Yang2017}.} In this section we will show that these results can be readily calculated by applying (\ref{PIRho4}) and (\ref{PIdensityProb}).

Denote the location of $T_1$ is $x_1$ and location of $T_2$ is $x_2$. Suppose both slits are opened. The electron passes through the double slit at $t_a$ and detected at the destination screen at $t_b$. \replaced{In this process, one only knows the electron interacts with the slits, but}{When the electron passes through the double slit, there is interaction between the electron and the slit. Denote the relational matrix after interaction as $R$. Suppose there is no entanglement after the electron passes through the slits,} there is no inference information on exactly which slit the electron passes through. Assuming equal probability for the electron passing either $T_1$ or $T_2$ at $t_a$, the state vector is represented by a superposition state $|\psi_a\rangle=(1/\sqrt{2})(|x_1\rangle + |x_2\rangle)$. The reduced density operator at $t_a$ is $\varrho(t_a)=(1/2)(|x_1\rangle + |x_2\rangle)(\langle x_1| + \langle x_2|)$. Its matrix element is
\begin{equation}
\label{splitRho}
\begin{split}
 \varrho(x_a, x'_a) & = \langle x_a|\hat{\rho}_a |x'_a\rangle \\
& =\frac{1}{2}(\delta(x_a-x_1) + \delta(x_a-x_2)) \\
& \times (\delta(x'_a-x_1) + \delta(x'_a-x_2))
\end{split}
\end{equation}
where the property $\langle x_a|x_i\rangle=\delta(x_a-x_i)$ is used in the last step. Later at time $t_b$, according to Eq.(\ref{PIRho4}), the density matrix element for the electron at a position $x_b$ in the detector screen is
\begin{equation}
\label{splitRho2}
\begin{split}
    & \varrho(x_b, x'_b)  = \\ & \frac{1}{2}\int\int_{-\infty}^{\infty} K(x_b, x_a)  \varrho(x_a, x'_a)K^*(x'_b, x'_a)dx_a dx'_a \\
    & = \frac{1}{2} [K(x_b,  x_1)K^*(x'_b, x_1)+ K(x_b, x_2)K^*(x'_b, x_2) \\
    & + K(x_b,  x_1)K^*(x'_b, x_2)+ K(x_b, x_2)K^*(x'_b, x_1)].
\end{split}
\end{equation}
According to (\ref{PIdensityProb}), the probability of finding the electron at a position $x_b$ is
\begin{equation}
\label{splitWF2}
\begin{split}
    p(x_b) & =  \varrho(x_b, x_b) \\
    & = \frac{1}{2} [K(x_b,  x_1)K^*(x_b, x_1)+ K(x_b, x_2)K^*(x_b, x_2) \\
    & + K(x_b,  x_1)K^*(x_b, x_2)+ K(x_b, x_2)K^*(x_b, x_1)] \\
    & = \frac{1}{2} |K(x_b,  x_1) + K(x_b, x_2)|^2\\
    & = \frac{1}{2} |\varphi_1(x_b) + \varphi_2(x_b)|^2
\end{split}
\end{equation}
where $\varphi_1(x_b)=K(x_b, x_1)$ and $\varphi_2(x_b)=K(x_b, x_2)$. Therefore, the probability to find the electron showing up at position $x_b$ is $p(x_b, t_b) = (1/2)|\varphi_1(x_b, t_b)+\varphi_2(x_b, t_b)|^2$. 

\begin{figure}
\begin{center}
\includegraphics[scale=3.0]{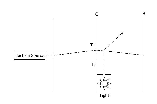}
\caption{\added{Modified double slit experiment discussed in page 7 of Ref.~\cite{Feynman05}. Here a light source is placed behind the double slit to assist detecting which slit an electron passes through.}}
\label{fig:2}       
\end{center}
\end{figure}

\replaced{Now consider the modified experiment proposed by Feynman, shown in Fig. 2. A lamp is placed behind the double slit, which enables one to detect which slit the electron passes through. The photon is scattered by the electron. By detecting the scattered photon one can infer which slit the electron passes through. In a sense, the electron is ``measured" by the photon. The variable carried by the photon to tell which slit the electron passed through acts as a pointer variable. There is a one to one correlation between the pointer state of the photon and which slit the electron passes through.}{Now suppose there is a weak measurement such that another apparatus, $A'$, is added to detect whether the electron passes through $T_1$ or $T_2$. $A'$ must interact with the electron in order to detect whether electron passes through $T_1$ or $T_2$.} Denote the pointer states of photon that are corresponding to the electron passing though $T_1$ and $T_2$ as $|T_1\rangle$ and $|T_2\rangle$, respectively. \deleted{After the interaction, there is an entanglement between the electron and $A'$. As a weak measurement (or, non-selective measurement), the quantum state of $S$ is not yet projected to either eigenstate of $|x_1\rangle$ or $|x_2\rangle$.} The state vector of the composite system of electron and photon \added{right after scattering} is $|\Psi\rangle=1/\sqrt{2}(|x_1\rangle|T_1\rangle + |x_2\rangle|T_2\rangle)$, thus the reduced density operator for the electron after passing the slit configuration is $\hat{\rho}_a  = Tr_T(|\Psi\rangle\langle\Psi|)=\frac{1}{2}(|x_1\rangle\langle x_1| + |x_2\rangle\langle x_2|)$. Its matrix element is 
\begin{equation}
\label{splitRho3}
\begin{split}
    & \varrho(x_a, x'_a, t_a) = \langle x_a|\hat{\rho}_a |x'_a\rangle \\
    &= \frac{1}{2}(\delta(x_a-x_1)\delta(x'_a-x_1) \\
    &+ \delta(x_a-x_2)\delta(x'_a-x_2)).
\end{split}
\end{equation}
Substituted this into Eq.(\ref{PIRho4}), the reduced density matrix element at $t_b$ is
\begin{equation}
\begin{split}
    & \varrho(x_b, x'_b) = \\ & \frac{1}{2}\int\int_{-\infty}^{\infty} K(x_b, x_a)  \varrho(x_a, x'_a)K^*(x'_b, x'_a)dx_a dx'_a \\
    & = \frac{1}{2} [K(x_b,  x_1)K^*(x'_b, x_1)+ K(x_b, x_2)K^*(x'_b, x_2)].
\end{split}
\end{equation}
From this, the probability of finding the electron at position $x_b$ is
\begin{equation}
\label{doubleSplitProb2}
\begin{split}
    p(x_b) & =  \varrho(x_b, x_b) \\
    & = \frac{1}{2} [K(x_b,  x_1)K^*(x_b, x_1)+ K(x_b, x_2)K^*(x_b, x_2)] \\
    & = \frac{1}{2} (|K(x_b,  x_1)|^2 + |K(x_b, x_2)|^2)\\
    & = \frac{1}{2} (|\varphi_1(x_b)|^2 + |\varphi_2(x_b)|^2)
\end{split}
\end{equation}
There is no interference term $\varphi_1\varphi_2$ in Eq.(\ref{doubleSplitProb2}). This result can be further understood as following. \replaced{The interaction between the electron and photon results in an entanglement between them. The alternatives of passing through $T1$ or $T2$ become distinguishable as it can be inferred from the correlation encoded in the photon. They are exclusive alternatives.}{The interaction between $S$ and $A'$ results in the entanglement between these two systems. This is equivalent to say that $A'$ has ``measured" $S$. Therefore, $A'$ has extract information from $S$. The indeterminacy of which eigenstate $S$ is in disappears since such information can be inferred if we know which eigenstate $A'$ is in.} Thus, the interference terms should be excluded to avoid over-counting alternatives that contribute to the measurement probability (see more detailed explanation of this probability counting rule in Ref.~\cite{Yang2017}), resulting in (\ref{doubleSplitProb2}). \added{These considerations are consistent with Feynman's concept of interfering alternatives versus exclusive alternative. Here we further advance the ideas by giving a concrete calculation of the resulting probability.}

\subsection{The Von Neumann Entropy}
\label{sec:entropy}
The definition of Von Neumann entropy in (\ref{vonNeumann2}) calls for taking the logarithm of the density matrix. This is a daunting computation task when the reduced density matrix element is defined using path integral formulation as Eq.(\ref{PIIF}). Brute force calculation of the eigenvalues of the reduced density matrix may be possible if one approximates the continuum of the position with a lattice model with spacing $\epsilon$, and then takes the lattice spacing $\epsilon\to 0$ to find the limit. On the other hand, in quantum field theory, there is a different approach to calculate entropy based on the ``replica trick"~\cite{Wilczek94, Cardy09, Takaya17}. This approach allows one to calculate the von Neumann entropy without taking the logarithm. We will briefly describe this approach and apply it here. In the case of finite degree of freedom, the eigenvalues of the reduced density matrix $\lambda_i$ must lie in the interval $[0,1]$ such that $\sum_i\lambda_i=1$. This means the sum $Tr(\rho^n)=\sum_i\lambda_i^n$ is convergent. This statement is true for any $n>1$ even $n$ is not an integer. Thus, the derivative of $Tr(\rho^n)$ with respect to $n$ exists. It can be shown that
\begin{equation}
    \label{entropy2}
    H(\rho_S) = -\lim_{n\to 1}\frac{\partial}{\partial n}Tr(\rho_S^n) = \lim_{n\to 1} H_S^{(n)}
\end{equation}
where $H_S^{(n)}$ is the R$\acute{e}$nyi entropy, defined as
\begin{equation}
    \label{Renyi}
    H_S^{(n)}=\frac{1}{1-n}lnTr(\rho_S^n).
\end{equation}
The replica trick calls for calculating $\rho_S^n$ for integers $n$ first and then analytically continuing to real number $n$. In this way, calculation of the von Neumann entropy is turned into the calculation of $Tr(\rho_S^n)$. But first, we have to construct $\rho_S^n$ from the path integral version of reduced density matrix element in Eq.(\ref{PIIF}). 

$\rho_S^n$ is basically the multiplication of $n$ copies of the same density matrix, i.e., $\rho_S^n = \rho_S^{(1)} \rho_S^{(2)} \ldots \rho_S^{(n)}$, at time $t_b$. To simplify the notation, we will drop the subscript $S$. Since all the calculations are at time $t_b$, we also drop subscript $b$ in Eq.(\ref{PIIF}). Denote the  element of $i$th copy of reduced density matrix as $\rho(x^{(i)}_-, x^{(i)}_+)$, where $x^{(i)}_- = x^{(i)}_L(t_b)$ and $x^{(i)}_+ = x^{(i)}_R(t_b)$ are two different positions at time $t_b$, and $x^{(i)}_L(t)$ and $x^{(i)}_R(t)$ are two different trajectories used to perform the path integral for $i$th copy of reduced density matrix element as defined in (\ref{PIIF}). With these notations, the matrix element of $\rho_S^n$ is
\begin{equation}
    \label{rhoN}
    \begin{split}
    &\rho^n(x^{(1)}_-, x^{(n)}_+) \\
    & =\int dx^{(1)}_+ dx^{(2)}_-\ldots dx^{(n-1)}_+ dx^{(n)}_- \rho(x^{(1)}_+, x^{(1)}_-)\ldots\\
    &\times \rho(x^{(n)}_+, x^{(n)}_-)\delta(x^{(1)}_+ - x^{(2)}_-)\ldots \delta(x^{(n-1)}_+ - x^{(n)}_-) \\
    & = \int\prod_{i=1}^{(n-1)}[dx^{(i)}_+dx^{(i+1)}_- \rho(x^{(i)}_-, x^{(i)}_+)  \\
    &\times \delta(x^{(i)}_+ - x^{(i+1)}_-)]\rho(x^{(n)}_-, x^{(n)}_+).
    \end{split}
\end{equation}
The delta function is introduced in the calculation above to ensure that when multiplying two matrices, the second index of the first matrix element and first index of the second matrix element are identical in the integral. From (\ref{rhoN}), we find the trace of $\rho^n$ is
\begin{equation}
    \label{rhoN2}
    \begin{split}
    & Tr(\rho^n)=\int \rho^n(x^{(1)}_-, x^{(n)}_+) \delta (x^{(n)}_+ - x^{(1)}_-) dx^{(n)}_+ dx^{(1)}_-\\
    & = \int\prod_{i=1}^{n}\{\rho(x^{(i)}_-, x^{(i)}_+) \delta(x^{(i)}_+ - x^{(mod(i, n)+1)}_-) \\
    &\times dx^{(i)}_+dx^{(mod(i,n)+1)}_-\}
    \end{split}
\end{equation}
where $mod(i, n)=i$ for $i<n$ and $mod(i, n)=0$ for $i=n$. To further simplifying the notation, denote $\Delta_{+-}^{(i)} = \delta(x^{(i)}_+ - x^{(mod(i, n)+1)}_-)$, $dx_{+-}^{(i)}=dx^{(i)}_+dx^{(mod(i,n)+1)}_-$ and $Z(n)=Tr(\rho^n)$, the above equation is rewritten in a more compact form
\begin{equation}
    \label{Zn}
    Z(n) = \int\prod_{i=1}^{n}\{\rho(x^{(i)}_-, x^{(i)}_+) \Delta_{+-}^{(i)} dx^{(i)}_{+-}\}
\end{equation}
Substituting Eq.(\ref{PIIF}) into (\ref{Zn}), we get
\begin{equation}
    \label{Zn2}
    \begin{split}
    Z(n) & = \int\prod_{i=1}^{n}\{\Delta_{+-}^{(i)} dx^{(i)}_{+-} \\ &\times \int_{{\cal{R}}_L^{(i)}}\int_{{\cal{R}}_R^{(i)}}e^{(i/\hbar)[S(x^{(i)}_L(t) - S(x^{(i)}_R(t))]} \\
    &\times F(x^{(i)}_L(t), x^{(i)}_R(t)){\cal{D}}x^{(i)}_L(t){\cal{D}}x^{(i)}_R(t) \}.
    \end{split}
\end{equation}
We have omitted the normalization so far. To remedy this, the normalized $Z(n)=Tr(\rho_n)$ should be rewritten as
\begin{equation}
    \label{NormalizedZn}
    {\cal{Z}}(n) = Tr(\varrho^n) = \frac{Z(n)}{Z(1)^n}.
\end{equation}
where $Z(1)=Z=Tr(\rho)$ as defined in Eq.(\ref{PIIF}). Once ${\cal{Z}}(n)$ is calculated, the von Neumann entropy is obtained through 
\begin{equation}
    \label{entropy3}
    H(\varrho) = -\lim_{n\to 1}\frac{\partial}{\partial n}Tr(\varrho_S^n) = -\lim_{n\to 1} \frac{\partial}{\partial n}{\cal{Z}}(n)
\end{equation}

(\ref{Zn2}) appears very complicated. Let's validate it in the case that there is no interaction between $S$ and $A$. One would expect there is no entanglement between $S$ and $A$ in this case. Thus, the entropy should be zero. We can check whether this is indeed the case based on (\ref{Zn2}). Since there is no interaction between $S$ and $A$, the influence functional is simply a constant. (\ref{PIIF}) is simplified into 
\begin{equation}
    \label{pureRho}
    \varrho(x_b, x'_b) = \frac{1}{Z} \varphi(x_b)\varphi^*(x'_b) = \frac{1}{Z}\rho(x_b, x'_b) 
\end{equation}
where $\varphi(x_b)$ has been given in Eq.(\ref{PIWF4}). Taking trace of the above equation, one gets
\begin{equation}
     1 = Tr(\varrho) = \frac{1}{Z} \int \varphi(x_b)\varphi^*(x_b)dx_b
\end{equation}
This gives
\begin{equation}
     Z = \int \varphi(x_b)\varphi^*(x_b)dx_b
\end{equation}
(\ref{pureRho}) implies that $S$ is in a pure state, since by definition, $\varphi(x_b) = \langle x_b|\varphi\rangle = \int\delta(x-x_b)\varphi(x)dx$, one can rewrite (\ref{pureRho}) in a form for a pure state,
\begin{equation}
     \varrho(x_b, x'_b) = \frac{1}{Z}\langle x_b|\varphi\rangle\langle\varphi| x'_b\rangle.
\end{equation}
Multiplication of density matrix that represents a pure state gives the same density matrix itself. Using the same notation as in (\ref{rhoN}), we obtain
\begin{equation}
    \label{rhoN4}
    \begin{split}
    &\rho^2(x^{(1)}_-, x^{(2)}_+) \\
     & = \int \rho(x^{(1)}_-, x^{(1)}_+)\rho(x^{(2)}_-, x^{(2)}_+)\delta(x^{(1)}_+ - x^{(2)}_-)dx^{(1)}_+dx^{(2)}_- \\
    & = \int \rho(x^{(1)}_-, x^{(1)}_+)\rho(x^{(1)}_+, x^{(2)}_+)dx^{(1)}_+ \\
     & = \int \varphi(x^{(1)}_-)\varphi^*( x^{(1)}_+)\varphi(x^{(1)}_+)\varphi^*( x^{(2)}_+)dx^{(1)}_+ \\
    & = Z\rho(x^{(1)}_-, x^{(2)}_+).
    \end{split}
\end{equation}
From this we can deduce that $\rho^n = Z^{n-1}\rho$. This gives $Z(n) = Tr(\rho^n) = Z^n$, and
\begin{equation}
    {\cal{Z}}(n) = \frac{Z(n)}{Z(1)^n} = 1.
\end{equation}
It is independent of $n$, thus
\begin{equation}
    \label{entropy4}
    H(\varrho) = -\lim_{n\to 1} \frac{\partial}{\partial n}{\cal{Z}}(n) = 0,
\end{equation}
as expected. The von Neumann entropy is only non-zero when there is an interaction between $S$ and $A$. The effect of the interaction is captured in the influence functional. Concrete form of the influence functional should be constructed in order to find examples where the von Neumann entropy is non-zero.

\section{Discussion and Conclusion}
\label{sec:conclusion}

\subsection{The $G$ Function}
The $G$ function introduced in Eq.(\ref{Gfunction}) can be rewritten in terms of the influence functional. To do this, we first rewrite Eq.(\ref{KFunction}) as
\begin{equation}
    \label{KFunction2}
    \begin{split}
    & K(x_a, x_b, z_b) = \int_{{\cal{R}}_{ab}^S}{\cal{D}}x(t)exp\{\frac{i}{\hbar}S^S_p(x(t)\}\\
    & \times \int_{{\cal{R}}_b^{A'}}{\cal{D}}z(t)exp\{\frac{i}{\hbar}[S^{A'}_p(z(t))+S^{SA'}_{int}(x(t), z(t))]\} 
    \end{split}
\end{equation}
Substituting this into Eq.(\ref{Gfunction}), we have
\begin{equation}
    \label{Gfunction2}
    \begin{split}
    G(x_a, x_b; x'_a, x'_b) &= \int_{{\cal{R}}_{ab}^S}\int_{{\cal{R}}_{a'b'}^S}e^{(i/\hbar)[S^S_p(x(t)) - S^S_p(x'(t))]} \\
    &\times F(x(t), x'(t)){\cal{D}}x(t){\cal{D}}x'(t)
    \end{split}
\end{equation}
Consequently, the normalized reduced density matrix element in Eq.(\ref{PIRho3}) becomes
\begin{equation}
    \label{PIRho5}
    \begin{split}
    \varrho((x_b, x'_b) &=\frac{1}{Z}
    \int\int_{{\cal{R}}_{ab}^S}\int_{{\cal{R}}_{a'b'}^S}e^{(i/\hbar)[S^S_p(x(t)) - S^S_p(x'(t))]} \\
    &\times \rho(x_a, x'_a)F(x(t), x'(t)){\cal{D}}x(t){\cal{D}}x'(t)dx_a dx'_a.
    \end{split}
\end{equation}
This gives the same result as in Ref.~\cite{FeynmanVernon}. However, there is advantage of using the $G$ function instead of the influence functional $F$ because the complexity of path integral is all captured inside the $G$ function, making it mathematically more convenient. This can be shown in the following modified double split experiment example. Suppose after the electron passing the double slit, there is \deleted{no detector next to the slits to detect which split the electron passing through, but there is another system $B$ that continues to interact with $S$ till $S$ reaches the destination screen detector. Alternatively, this can be} an electromagnetic field between the double slit and the destination screen. We will need to apply (\ref{PIRho3}) instead of (\ref{PIRho4}) to calculate the reduced density matrix element. In this case, we simply substitute (\ref{splitRho}) into (\ref{PIRho3}) and obtain
\begin{equation}
\begin{split}
    \varrho(x_b, x'_b) & = \frac{1}{2} [G(x_1, x_b; x_1, x'_b) + G(x_1, x_b; x_2, x'_b) \\
    & + G(x_2, x_b; x_1, x'_b) + G(x_2, x_b; x_2, x'_b)].
\end{split}
\end{equation}
The probability of finding the electron at position $x_b$ is
\begin{equation}
\label{doubleSplitProb22}
\begin{split}
    p(x_b) & =  \varrho(x_b, x_b) \\
    & = \frac{1}{2} [G(x_1, x_b; x_1, x_b) + G(x_1, x_b; x_2, x_b) \\
    & + G(x_2, x_b; x_1, x_b) + G(x_2, x_b; x_2, x_b)].
\end{split}
\end{equation}
From the definition of $G(x_a, x_b; x'_a, x'_b)$ in (\ref{Gfunction}), it is easy to derive the following property,
\begin{equation}
    \label{Gproperty1}
    G(x_a, x_b; x'_a, x'_b) = G^*(x'_a, x'_b; x_a, x_b).
\end{equation}
When $x_a=x'_a$ and $x_b=x'_b$, we have $G(x_a, x_b; x_a, x_b) = G^*(x_a, x_b; x_a, x_b)$. Thus, $G(x_a, x_b; x_a, x_b)$ must be a real function. We denote it as $G_R(x_a, x_b)$. With these properties, Eq.(\ref{doubleSplitProb22}) can be further simplified as
\begin{equation}
\label{doubleSplitProb3}
\begin{split}
    p(x_b) & = \frac{1}{2} [G_R(x_1, x_b) + G_R(x_2, x_b)] \\
    & + Re[G(x_1, x_b; x_2, x_b)].
\end{split}
\end{equation}
This is consistent with the requirement that $p(x_b)$ must be a real number. The second term $Re[G(x_1, x_b; x_2, x_b)]$ is an interference quantum effect due to the fact that the initial state after passing through the double slit is a pure state. This interference term also depends on the interaction between the electron and the electromagnetic field. If the electromagnetic field is adjustable, the probability distribution will be adjusted accordingly. Tuning the electromagnetic field will cause the probability distribution $p(x_b)$ to change through the interference term in (\ref{doubleSplitProb3}). Presumably, the Aharonov$-$Bohm effect~\cite{Bohm59} can be explained through (\ref{doubleSplitProb3}) as well. The detailed calculation will be reported in a separated manuscript.

\subsection{Influence Functional and Entanglement Entropy}
\label{subsec:entanglement2}
In Section \ref{sec:entropy}, we show that when there is no interaction, the influence functional is a constant and therefore the entanglement entropy is zero. We can relax this condition to include interaction but want to detect whether there is entanglement. Suppose the influence functional can be decomposed in the following production form,
\begin{equation}
    \label{IFcondition}
    F(x(t), x'(t)) = f(x(t))f^*(x'(t)).
\end{equation}
Such a form of influence functional satisfies the rule~\cite{Feynman05} $F(x(t), x'(t)) = F^*(x'(t), x(t))$. Eq.(\ref{IFcondition}) implies that the entanglement entropy is still zero even there is interaction between $S$ and $A$. The reason for this is that the reduced density matrix element can be still written as the form of Eq.(\ref{pureRho}) but with 
\begin{equation}
    \label{PIWF9}
    \varphi(x_b) =\int_{{\cal{R}}_b} e^{(i/\hbar)S(x(t))}f(x(t)){\cal{D}}x(t).
\end{equation}
As long as (\ref{pureRho}) is valid, $S$ is in a pure state therefore the reasoning process from (\ref{pureRho}) to (\ref{entropy4}) is applicable here.

We now examine the entanglement entropy for some general forms of influence functional. The most general exponential functional in linear form is~\cite{Feynman05}
\begin{equation}
    \label{LinearIF}
    F(x(t), x'(t)) = exp\{i\int x(t)g(t)dt - i\int x'(t)g(t)dt]\}
\end{equation}
where $g(t)$ is a real function. Clearly this form satisfies the condition specified in Eq.(\ref{IFcondition}) since we can take $f(x(t)) = exp[i\int x(t)g(t)dt]$ and $f(x'(t)) = exp[i\int x'(t)g(t)dt]$. The entanglement entropy is zero with this form of influence functional.

On the other hand, the most general Gaussian influence functional~\cite{Feynman05}
\begin{equation}
    \label{GaussianIF}
    \begin{split}
    F(x(t), x'(t))& = exp\{\int\int [\alpha(t, t')x(t') \\
    & - \alpha^*(t, t')x'(t')][x(t)-x'(t)]dtdt'\}
    \end{split}
\end{equation}
where $\alpha(t, t')$ is an arbitrary complex function, defined only for $t>t'$. This form of influence functional cannot be decomposed to satisfy the condition specified in Eq.(\ref{IFcondition}). The entanglement entropy may not be zero in this case. It will be of interest to further study the influence functional of some actual physical setup and calculate the entanglement entropy explicitly.

\subsection{\added{Compared to the Transactional Interpretation}}
\label{subsec:transaction}
\added{After the initial relational formulation of quantum mechanics~\cite{Yang2017}, it was brought to our attention that the bidirectional measurement process which is important in the derivation of the measurement probability appears sharing some common ideas with the transaction model~\cite{Cramer}. In particular, the transaction model requires a handshake between a retarded ``offered wave" from an emitter and an advanced ``confirmation wave" from an absorber to complete a transaction in a quantum event. This is a bidirectional process. While it is encouraging to note that the bidirectional nature of a quantum event has been recognized in the transaction model, there are several fundamental differences between the transaction model and the bidirectional measurement framework presented here. First, the transaction model considers the offered wave and the confirmation wave as real physical waves. In our framework, we do not assume such waves existing. Instead, the probing and responding are just two aspects in a measurement event, and we require the probability calculation to faithfully model such bidirectional process. Second, in the transaction model, the randomness of measurement outcomes is due to the existence of different potential future absorbers. Thus, the randomness in quantum mechanics depends on the existence of absorbers. There is no such assumption in our framework. Third, the transaction model derives that the amplitude of the confirmation wave at the emitter is proportional to the modulus square of the amplitude of the offer wave, which is related to the probability of completing a transaction with the absorber. This appears to be ambiguous since it suggests the confirmation wave is also a probability wave. In our formulation, we only focus on how the measurement probability is calculated, and clearly point out that the wave function is just a mathematical tool.}

\subsection{\added{Outlook of the Relational Formulation}}
\label{sec:outlook}
\added{The framework to calculate the measurement probability in Section \ref{sub:measProb} is the key to our reformulation. However, it is essentially based on an operational model from a detailed analysis of bidirectional measurement process, instead of being derived from more fundamental physical principles. In particular, there may be better justification for Eq. (\ref{conjugate}). The current model is only served as a step to deepen our understanding of relational quantum mechanics. It is desirable to continue searching for more fundamental physical principles to justify the calculation of measurement probability. The fact the $R^{SA}_{ij}$ is a complex number means that this variable actually comprises two independent variables, the modulus and the phase. This implies that more degrees of freedom are needed to have a complete description of a quantum event. Stochastic mechanics, for instance, introduces forward and backward velocities instead of just one classical velocity to describe the stochastic dynamics of a point particle. With the additional degree of freedom, two stochastic differential equations for the two velocities are derived. Then, through a mathematical transformation of two velocity variables in $\mathbb{R}$ into one complex variable in $\mathbb{C}$, the two differential equations turn into the Schr\"{o}dinger equation~\cite{Nelson, Nelsonbook, Yasue, Guerra, Yang2021}. It will be interesting to investigate if $R^{SA}_{ij}$ can be implemented in the context of stochastic mechanics, where we expect $R^{SA}_{ij}$ will be decomposed to two independent variables in $\mathbb{R}$ and each of them is a function of velocity variables.

As discussed in the introduction section, the RQM principle consists two aspects. First, we need to reformulate quantum mechanics relative to a QRF which can be in a superposition quantum state, and show how quantum theory is transformed when switching QRFs. In this aspect, we accept the basic quantum theory as it is, including Schr\"{o}dinger equation, Born's rule, von Neunman measurement theory, but add the QRF into the formulations and derive the transformation theory when switching QRFs~\cite{Brukner, Hoehn2018, Yang2020}. Second, we go deeper to reformulate the basic theory itself from \textit{relational properties}, but relative to a fixed QRF. Here the fixed QRF is assume to be in a simple eigen state. This is what we do in Ref.~\cite{Yang2017, Yang2017_2} and the present work. A complete RQM theory should combine these two aspects together. This means one will need to reformulate the basic quantum theory from relational properties and relative to a quantum reference frame that exhibits superposition behavior. Therefore, a future step is to investigate how the relational probability amplitude matrix should be formulated when the QRF possesses superposition properties, and how the relational probability amplitude matrix is transformed when switching QRFs.}

\subsection{Conclusions}
\replaced{The full implementation of probability amplitude using path integral provides a concrete physical picture for the relational formulation of quantum mechanics, which is missing in the initial formulation~\cite{Yang2017}. This gives a clearer meaning of the relational probability amplitude. $R^{SA}_{ij}$ is defined as the summation of quantity $e^{iS_p/\hbar}$, where $S_p$ is the action of the composite system $S+A$ along a path. Physical interaction between $S$ and $A$ may cause change of $S_p$, which is the phase of a complex number. Such definition is consistent with the analysis in Section \ref{sub:measProb} on the factors that determine the relational probability amplitude. In return, the implementation brings back new insight to the path integral itself by completing the justification on why the measurement probability can be calculated as modulus square of probability amplitude.

The path integral implementation allows us to develop formulations for some of interesting physical processes and concepts. For instance, we can describe the interaction history between the measured system and a series of measuring systems or environment. This confirms the idea that a quantum state essentially encodes the information of previous interaction history~\cite{Rovelli07}. A more interesting application of the coordinator representation of the reduced density matrix is the method to calculate entanglement entropy using path integral approach. This will allow us to potentially calculate entanglement entropy of a physical system that interacts with classical fields, such as an electron in an electromagnetic field. Section \ref{subsec:entanglement2} gives a criterion on whether there is entanglement between the system and external environment based on the influence functional.

The present work significantly extends the initial formulation of RQM~\cite{Yang2017} in several fronts. In addition to the concrete path integral implementation of relational probability amplitude, we also provide a much clearer explanation on the framework for calculating the measurement probability from the bidirectional measurement process, as discussed in Section \ref{sub:measProb}. At the conceptual level, a thorough analysis of the two aspects of relational quantum mechanics connects this work with the QRFs theories, which are currently under active investigations. As a result, one can conceive the next step for a full relational formulation, as suggested in Section \ref{sec:outlook}.

Ref.~\cite{Yang2017, Yang2017_2} and this paper together show that quantum mechanics can be constructed by shifting the starting point from the independent properties of a quantum system to the relational properties between a measured quantum system and a measuring quantum system. In essence, quantum mechanics demands a new set of rules to calculate probability of a potential outcome from the physical interaction in a quantum measurement. Although the bidirectional measurement process is still an operational model, rather than deriving from first principles, we hope the reformulation efforts can be one step towards a better understanding of quantum mechanics.}
{We show that the relational probability amplitude can be implemented using path integral approach. The formulation is consistent with the results from the traditional path integral quantum mechanics. The significance of such implementation is two-folds. First, it gives a clearer meaning of the relational probability amplitude. $R^{SA}_{ij}$ is defined as the sum of quantity $e^{iS_p/\hbar}$, where $S_p$ is the action of the composite system $S+A$ along a path. Physical interaction between $S$ and $A$ may cause change of $S_p$, which is the phase of a complex number. This complex number, $e^{iS_p/\hbar}$, is just a probabilistic quantity, instead of a quantity associated with a physical property. Second, it gives a natural derivation of the coordinator representation of the reduced density matrix of the observed system. Based on the coordinator representation of the reduced density matrix element, it is mathematically convenient to develop formulation for some of interesting physical processes and concepts. For instance, we can describe the interaction history of the measured system and a series of measuring systems or environment, and use this formulation to elegantly explain the double slit experiment. A more interesting application of the coordinator representation of the reduced density matrix is the method to calculate entanglement entropy using path integral approach. This will allow us to potentially calculate entanglement entropy of a physical system that interacts with classical fields, such as an electron in an electromagnetic field. Section \ref{subsec:entanglement2} gives a criterion on whether there is entanglement between the system and external environment based on the influence functional. Since entanglement entropy is an important concept in quantum information theory, the method described here may lead to new insight on the information aspect of a physical system that interacts with classical fields. This is a topic for further research.

Ref.~\cite{Yang2017, Yang2017_2} and this paper together show that quantum mechanics can be constructed by shifting the starting point from the independent properties of a quantum system to the relational properties between a measured quantum system and a measuring quantum system. In essence, quantum mechanics demands a new set of rules to calculate probability of a potential outcome from the physical interaction in quantum measurement. The relational formulation of quantum mechanics is not only significant at the conceptual level, but also valuable at the application level. At the conceptual level, the difference between the relational formulation and traditional formulation results in fundamental consequence in some special scenario. This is demonstrated in resolving the EPR paradox~\cite{Yang2017_2}. At the application level, the path integral implementation of relational probability amplitude leads to new tool to calculate entanglement entropy based on the coordinator representation of the reduced density matrix. Ultimately, we hope the relational formulation presented in Ref.~\cite{Yang2017, Yang2017_2} and this paper can be one step towards a better understanding of quantum mechanics.}



%
%

\begin{acknowledgements}
\added{The author would like to thank the anonymous referees for their valuable comments, which motivate the author to improve the explanations of the measurement probability framework, and to clarify the relevancy of this work with the quantum reference frame theories.}
\end{acknowledgements}






\appendix

\end{document}